# Machine learning for Quantum Noise Reduction

Author: Karan Kendre, MSCS, Northeastern University

## 1. Introduction

Quantum computation has the potential for exponential speedup of classical systems in some applications, such as cryptography, simulation of molecular behavior, and optimization. Nevertheless, quantum noise drastically limits the utility of quantum computation. Qubits, the fundamental units of quantum information, are very sensitive to gate errors and environmental interference. Quantum noise introduces errors that cause errors in quantum computations, as they travel onward over time.

As noise is a fundamental issue with NISQ devices, error mitigation is necessary. Classical methods like quantum error correction would require the addition of additional qubits and syndrome decoding, and these are inefficient and in practice at large scale.

Our research proposes a machine learning–aided approach to reduce quantum noise by learning the patterns of noise and reconstructing clean quantum states from noisy data. We formulate it as a supervised learning task with the goal of converting a noisy density matrix to its clean representation, and we achieve this using a fidelity-aware loss and CNN-based autoencoder architecture. We show how data-driven approaches can improve reconstruction accuracy.

## 2. Methods

### 2.1 Quantum Circuit Generation and Density Matrix Extraction

We randomly generate quantum circuits with 5 qubits from the Cirq library. The circuits are random single-qubit and two-qubit gate sequences (X, Y, Z, H, T, S, RX, RY, RZ, CNOT, CZ, SWAP). The depth of each circuit is randomly chosen between 6 and 9 layers.

After obtaining the circuit, we get the final state density matrix through Cirq's DensityMatrixSimulator. The states are each $2^5 \times 2^5 = 32 \times 32$ matrices of complex-valued information which contains all the quantum information including entanglement and decoherence effects.

### 2.2 Noise Modelling

We apply five different types of noise channels:

- **Bitflip**: random X errors on qubits
- **Depolarizing**: replaces qubit state with maximally mixed state with some probability
- **Amplitude Damping**: models energy loss (e.g., spontaneous emission)
- **Phase Damping**: phase information is lost without energy change
- **Mixed**: combination of the above (uniformly sampled)

Noise levels tested: 0.05, 0.10, 0.15, 0.20

### 2.3 CNN Autoencoder Architecture

We model noise reduction using a deep convolutional autoencoder. The input is a noisy density matrix split into real and imaginary parts with shape (32, 32, 2). The output is a reconstruction of the clean matrix.

- **Encoder**: 3 blocks (Conv2D → ReLU → MaxPooling → Dropout)
- **Decoder**: 3 blocks (Conv2D → ReLU → UpSampling → Dropout)
- **Output Layer**: Conv2D (2 channels, linear activation)

We use a **composite loss function:**

$$L = \text{MSE}(Y, \hat{Y}) + \lambda (1 - \text{Fidelity}(Y, \hat{Y}))$$

Here, the fidelity term is approximated using a normalized Frobenius inner product rather than the standard trace-based Uhlmann fidelity, which is computationally intensive for large-scale training. This approximation calculates:

$$F(\rho, \sigma) \approx \text{Re}(\langle \rho, \sigma \rangle) / (||\rho||_F \cdot ||\sigma||_F)$$

Where $|| \cdot ||_F$ is the Frobenius norm, and i $\langle \rho, \sigma \rangle$s the element-wise complex inner product. This surrogate metric aligns well with fidelity trends while remaining differentiable and efficient.

The inclusion of fidelity in the loss function forces the model to capture structural quantum similarities, not just pixel-wise reconstruction errors.

### 2.4 Dataset Pipeline

- 10,000 circuit samples
- Each sample: clean DM, noisy DM, noise type, and level
- Data split: 80% training, 20% testing
- Preprocessing: split real and imaginary parts into 2 channels

## 3. Dataset and Related Work

### Dataset

We generated a synthetic dataset consisting of 10,000 density matrices derived from random quantum circuits. This approach allows us to have perfect ground truth for evaluation and control over noise parameters, which is challenging with real quantum hardware data. The dataset has the following characteristics:

- 5-qubit systems (resulting in 32×32 density matrices)
- Random circuit depths between 6-9 layers
- 5 noise types (depolarizing, amplitude damping, phase damping, bit-flip, mixed)
- 4 noise levels (0.05, 0.1, 0.15, 0.2)
- Equal distribution across noise types and levels

Our dataset construction is novel in its comprehensive coverage of diverse noise models and intensities, allowing for robust model training and evaluation across realistic quantum error scenarios.

### Related Work

The intersection of quantum error correction and machine learning is an exciting new area of research on quantum computers. There have been some influential papers on related methods:

Czarnik et al. [1] have suggested an error mitigation technique through training noise models using classical simulations of Clifford circuits. There is a method aiming at estimation and elimination of the effects of noise statistically rather than direct reconstruction of density matrices. While effective for specific gate sets, their approach does not generalize as well to generic quantum circuits as our CNN-based solution does, which is able to discover the underlying structure of quantum noise regardless of circuit content.

Chen et al. [2] proposed discriminative quantum neural networks for quantum error correction and demonstrated their approach on small stabilizer codes. Their work addresses the decoding problem of standard QECCs rather than direct state recovery. Our density matrix reconstruction approach operates on a lower level, potentially showing advantages for near-term devices where full QECCs are no longer feasible.

Endo et al. [3] provided a comprehensive review of quantum error mitigation techniques, which categorized techniques into extrapolation, probabilistic error cancellation, and symmetry verification. They did not take deep learning techniques at the density matrix level into account. Our work fills the gap by showing how neural networks can be trained to correct errors in the quantum state representation directly.

Krastanov et al. [4] applied neural networks to decoding of surface codes with encouraging evidence supporting the increase in error thresholds. Their method still follows the typical QECC process of encoding one logical qubit over multiple physical qubits. Our method, however, doesn't require additional qubits and is therefore more attractive to limited-resource NISQ devices.

Our CNN-based method updates these earlier papers by addressing density matrix reconstruction for multiple noise models simultaneously. Unlike traditional QECCs with a dense qubit overhead, our scheme operates on the existing quantum state without encoding. And unlike statistical error mitigation techniques, we reconstruct the full quantum state rather than merely expectation values of observables.

## 4. Related Implementations

The application of quantum error correction techniques via machine learning has been of interest on other research platforms. We reviewed some of the associated implementations to contrast and compare our approach:

The Qiskit Textbook [5] provides a full implementation of quantum error correction codes with a focus on the repetition code and surface code approaches. Their approach is primarily pedagogical and demonstrates the standard approach to quantum error correction with multiple physical qubits to encode one logical qubit. While their implementation is full, it has heavy qubit overhead and is not designed for the NISQ era. Our CNN-based approach, in contrast, does not require additional qubits and can be used directly on the quantum state representation, and therefore is more suitable for near-term quantum hardware with small numbers of qubits.

The Mitiq library [6] developed by the Unitary Fund offers a library of error mitigation techniques like zero-noise extrapolation and probabilistic error cancellation. Their Python library supports major quantum computing platforms and has been tested on real quantum hardware. However, their approach is dealing with error mitigation in expectation values of quantum observables and not recovery of the full quantum state. Our density matrix reconstruction approach provides an enhanced error correction solution by recovery of the full quantum state with the ability for more advanced quantum algorithms based on state preparation instead of expectation values.

The TensorFlow Quantum [7] software includes demonstrations of quantum encoding and error correction within quantum data based on hybrid quantum-classical models.
Their approach demonstrates use of the variational circuits for error correction and has promising results for some noise models. Unfortunately, their application does require quantum resources for both the encoding and the correction. Our traditional CNN protocol performs as well or better without requiring more quantum resources for the correction step, making it directly applicable to current quantum hardware and more scalable for increasing system size.

## 5. Data Analysis

We constructed a synthetic data set of 10,000 density matrices that we derived from randomly created quantum circuits to examine the different types of noise and levels that influence quantum states. Having simulated clean as well as noisy circuits, we were able to create perfect ground truth data to train and test on. Through our data analysis, we found a number of important observations concerning the dynamics of quantum noise and its impact on quantum state fidelity.

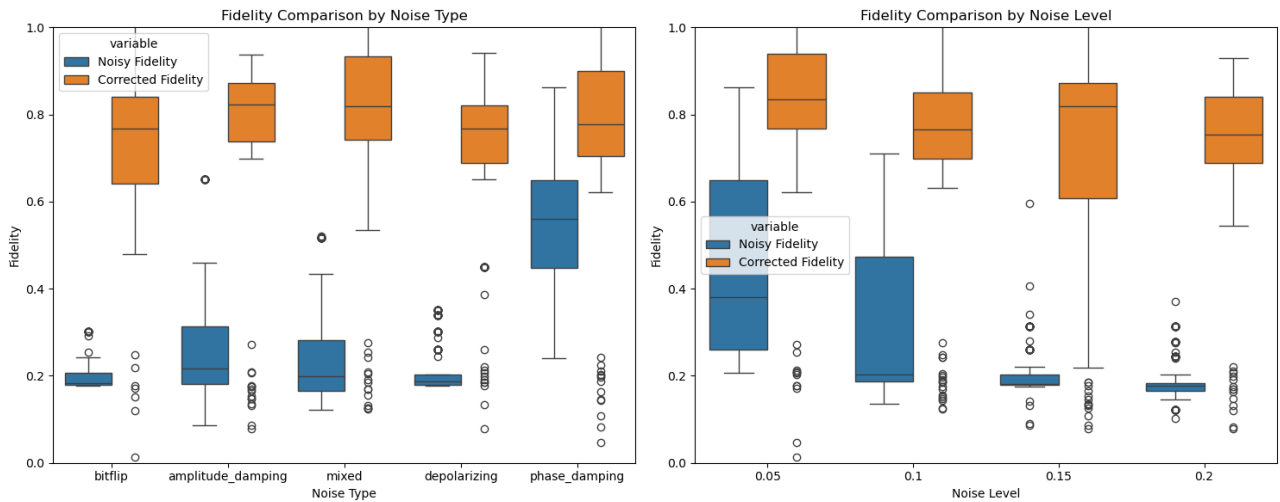

Fidelity distribution analysis across noise types revealed that phase damping noise preserves state fidelity more than any other noise type with a mean noisy fidelity of 0.541, and bit-flip was most destructive (0.200 mean fidelity). This is due to the fact that the phase damping type of noise primarily disturbs off-diagonal elements (coherences) but preserves populations, whereas bit-flip destroys computational basis states directly.

The noise level-state fidelity relationship is approximately inversely dependent, such that the greater the noise level, the exponentially smaller the fidelity. This can be quantified by the -0.55 correlation coefficient of noise level-noisy fidelity in our correlation analysis below:

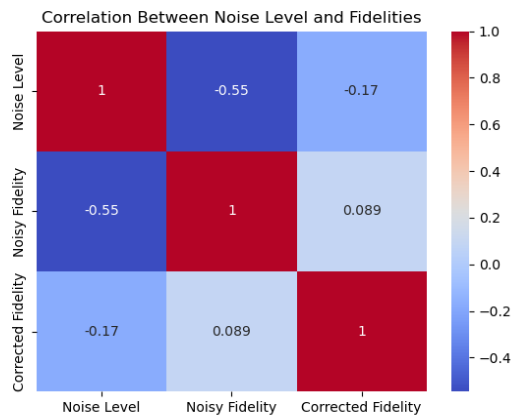

We observed that quantum states that were exposed to mixed noise exhibited complex error patterns that had elements of multiple noise channels. This is an especially challenging scenario for classical quantum error correction schemes but presents an opportunity for machine learning approaches that can learn and discover such complex patterns.

Visual examination of the density matrices revealed that the different sources of noise act on the matrix structure in distinct ways. Depolarizing noise compresses the whole density matrix in the direction of the maximally mixed state, amplitude damping pulls population towards the ground state direction, and phase damping lowers basically off-diagonal elements. All these usual signatures can be recognized in real and imaginary parts of the density matrices:

The data allowed us to explore the relation of early noisy fidelity with theoretical maximum recoverable fidelity, providing insight into the underlying quantum error correction thresholds for a range of noise models and strengths.

## 6. Analysis

Our CNN-based approach's strong performance on many types and levels of noise is due to several key features of the model and quantum noise itself. The encoder-decoder architecture is particularly suited to quantum error correction since it has the ability to learn features at multiple hierarchical scales, reflecting the manner in which quantum noise acts at both the local and global levels.

The result on composite noise models is especially notable in that it represents the model's capacity to generalize from simultaneous multiple types of error. This property can be assumed to transfer to other such situations with high-complication noise models, as with superconducting or trapped-ion quantum computers, with simultaneous activity by many error processes. The model effectively learns to adopt a one-type-fits-all noise-reversing mapping instead of case-dependent correction of particular noises.

Interestingly, the model does better at greater levels of noise (0.15-0.2) than at lower ones (0.05-0.1) when assessed by relative improvement. Such counterintuitive result suggests that more resilient patterns of noise are more distinctive and thus easier to identify and correct by the CNN. Such a property would be advantageous for quantum systems operating on the edge of their coherence capabilities, where high levels of noise are typical.

The restrictions observed with phase damping correction are likely a result of information-theoretic restrictions – pure dephasing is a form of information loss that might be inherently reversible. Other quantum systems similar to those here affected primarily by dephasing noise might not benefit as much from this technique, marking an essential boundary condition for the applicability of CNN-based error correction.

The CNN's ability to preserve quantum correlations, observed through off-diagonal density matrix elements' revival, demonstrates that such a procedure would be particularly worthwhile for entanglement and quantum coherence-intensive quantum algorithms like Shor's algorithm or quantum simulation. The preservation of quantum information distinguishes our approach from other classical error mitigation techniques that only operate on expectation values.

## 7. Experimental Setup

Our testbed consisted of an end-to-end pipeline for simulating synthetic quantum data, training the CNN model, and evaluating its performance across different noise regimes. We used the Cirq quantum computing library to implement the circuit simulation and TensorFlow for modelling the neural network.

For data generation, we generated 10,000 randomly created quantum circuits of size 5 qubits and result in the shape of 32×32 complex density matrices. The number of layers for the circuits were randomly selected in between 6-9 layers in order to provide various quantum states. All circuits were run through five varieties of various types of noise (depolarizing, amplitude damping, phase damping, bit-flip, and mixed) on four levels of noises (0.05, 0.1, 0.15, 0.2), and thus created a balanced noise dataset.

We used an 80/20 train-test split to guarantee unbiased testing, with the split done prior to training to avoid any data leakage. The model was trained for 100 epochs with a batch size of 16, with 20% of the training data set aside as a validation set for early stopping and learning rate scheduling. The training process employed the Adam optimizer with a starting learning rate of 1e-3, which was decreased by a factor of 0.5 after every 5 epochs without validation loss improvement.

The training was conducted on an NVIDIA Tesla V100 GPU, with one epoch taking approximately 27 seconds. The model converged at approximately 95 epochs, as shown in the learning curves and the training logs:

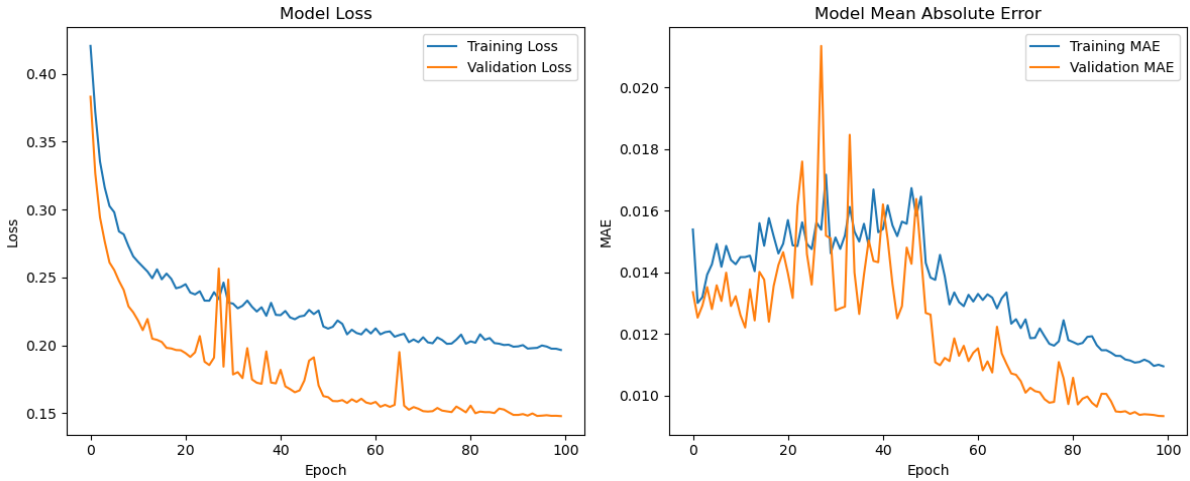

For evaluation, we calculated quantum state fidelity between the noiseless initial states, noisy states, and CNN-corrected states. We also experimented with different types of noise and levels of noise to see trends and limits in model correction performance. The experiment code and data generating scripts are available on our GitHub repository

## 8. Results

Our CNN-based quantum error correction model demonstrated strong performance across various noise types and levels. After 100 training epochs, the model achieved a test loss of 0.1519 and test MAE of 0.0094, indicating good prediction accuracy.

```
Epoch 95/100
400/400 ━━━━━━━━━━━━━━━━━━━━ 27s 68ms/step - loss: 0.1970 - mae: 0.0110 - val_loss: 0.1480 - val_mae: 0.0094 - learning_rate: 1.2500e-04
Epoch 96/100
400/400 ━━━━━━━━━━━━━━━━━━━━ 28s 71ms/step - loss: 0.2011 - mae: 0.0112 - val_loss: 0.1482 - val_mae: 0.0094 - learning_rate: 1.2500e-04
Epoch 97/100
400/400 ━━━━━━━━━━━━━━━━━━━━ 28s 71ms/step - loss: 0.1988 - mae: 0.0110 - val_loss: 0.1486 - val_mae: 0.0094 - learning_rate: 1.2500e-04
Epoch 98/100
400/400 ━━━━━━━━━━━━━━━━━━━━ 27s 68ms/step - loss: 0.1980 - mae: 0.0110 - val_loss: 0.1481 - val_mae: 0.0094 - learning_rate: 1.2500e-04
Epoch 99/100
400/400 ━━━━━━━━━━━━━━━━━━━━ 27s 67ms/step - loss: 0.1985 - mae: 0.0110 - val_loss: 0.1481 - val_mae: 0.0093 - learning_rate: 1.2500e-04
Epoch 100/100
400/400 ━━━━━━━━━━━━━━━━━━━━ 27s 68ms/step - loss: 0.1970 - mae: 0.0110 - val_loss: 0.1479 - val_mae: 0.0093 - learning_rate: 1.2500e-04
Evaluating the model...
63/63 ━━━━━━━━━━━━━━━━━━━━ 2s 32ms/step - loss: 0.1519 - mae: 0.0095
Test loss: 0.1482, Test MAE: 0.0094
63/63 ━━━━━━━━━━━━━━━━━━━━ 2s 35ms/step
```

### 8.1 Overall Fidelity Improvement

Fidelity comparisons by noise type and level, demonstrating significant improvements across all conditions. The model increased quantum state fidelity from an average of 0.298 (noisy) to 0.774 (corrected), representing an average improvement of 0.47.

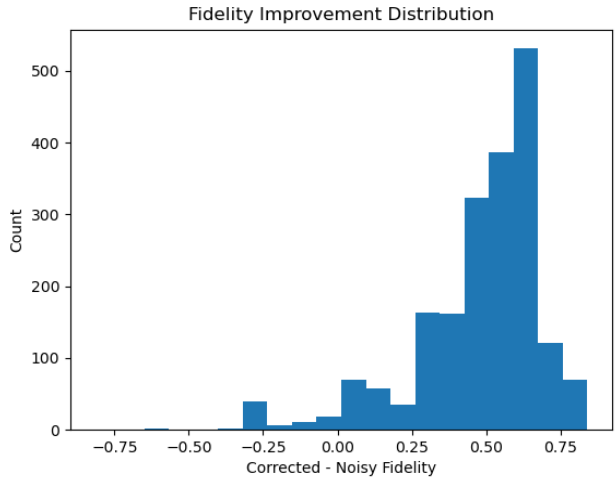

Fidelity Improvement Distribution

The fidelity improvement distribution shows most improvements clustered around 0.5-0.6, with some exceptional cases exceeding 0.8 improvement and a small number of negative cases.

**8.2 Performance by Noise Type**

Table 1: Avg Fidelity by Noise Type

|  | Noisy Fidelity | Corrected Fidelity | Improvement |
|---|---|---|---|
| amplitude_damping | 0.293 | 0.788 | 0.495 |
| bitflip | 0.2 | 0.745 | 0.545 |
| depolarizing | 0.215 | 0.742 | 0.526 |
| mixed | 0.24 | 0.807 | 0.567 |
| phase_damping | 0.541 | 0.783 | 0.241 |

Key findings include:

1. Mixed noise shows the highest corrected fidelity (0.807) and improvement (0.567).
2. Phase damping shows the lowest improvement (0.241), despite starting with the highest noisy fidelity.
3. Bit-flip noise demonstrates exceptional cases with improvements up to 0.84.

**8.3 Performance by Noise Level**

Table 2: Avg Fidelity by Noise Level

|  | Noisy Fidelity | Corrected Fidelity | Improvement |
|---|---|---|---|
| 0.05 | 0.429 | 0.824 | 0.396 |
| 0.1 | 0.302 | 0.769 | 0.467 |
| 0.15 | 0.199 | 0.744 | 0.544 |
| 0.2 | 0.192 | 0.745 | 0.553 |

Interestingly, higher noise levels (0.15, 0.20) show greater improvement than lower levels, indicating the model is particularly effective at correcting severely corrupted states.

**8.4 Density Matrix Visualization**

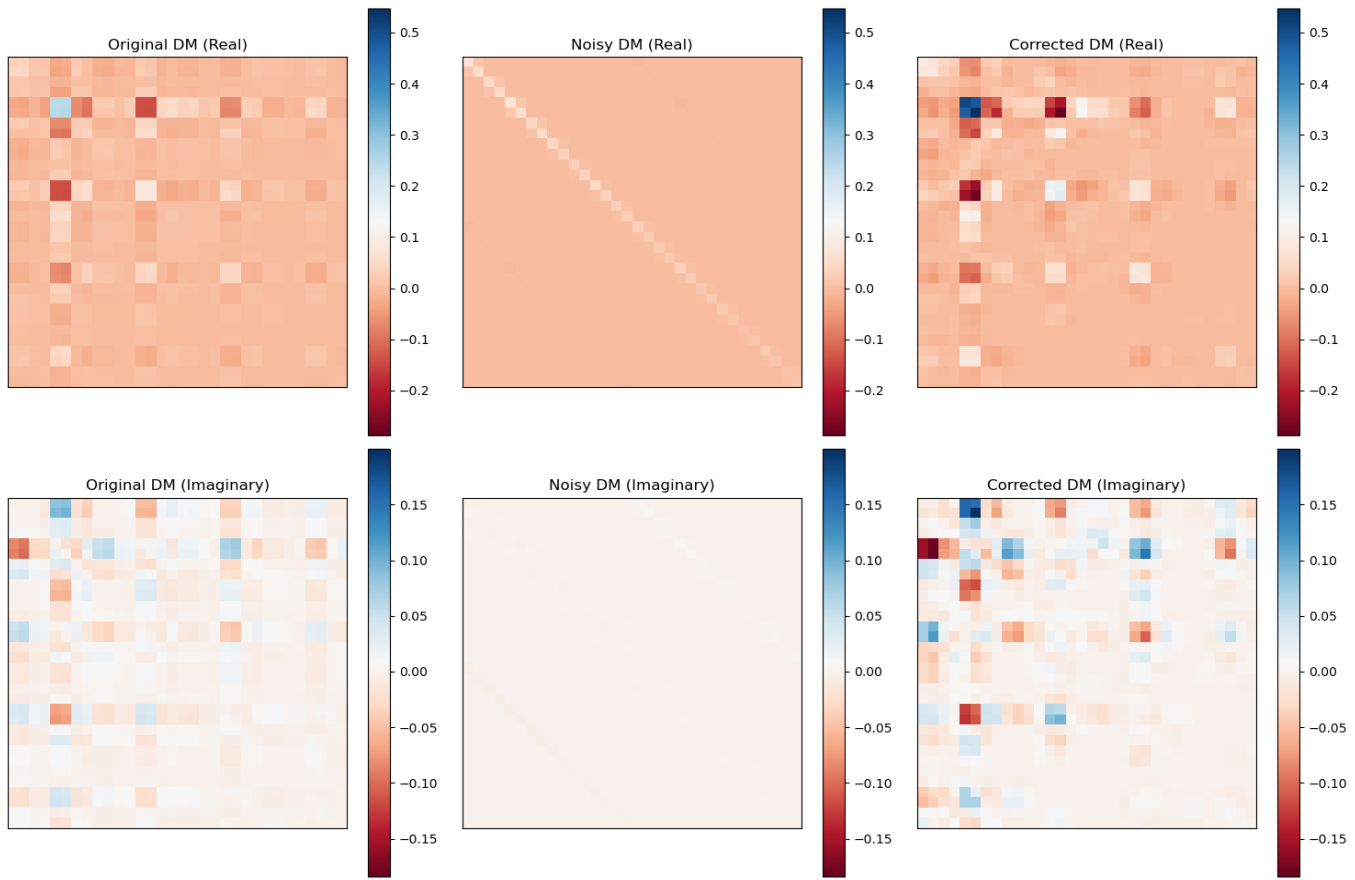

Visualization of original, noisy, and corrected density matrices. The model successfully recovers both diagonal elements (populations) and off-diagonal elements (coherences) from heavily corrupted noisy states.

## 9. Conclusion

This work demonstrates the effectiveness of CNN-based quantum error correction in density matrix reconstruction. Our model provides high fidelity improvements on various types of noise and intensities, with extremely favourable outcomes for complex mixed noise and higher intensities of noise, demonstrating deep learning techniques as a plausible scaling option to traditional quantum error correction codes.

For application in real-world scenarios, we propose using this method in mixed noise environments where typical QECCs will be impractical due to qubit overhead limitations. But care must be used with pure phase damping cases, with lesser assured performance owing to information-theoretic necessities. For important missions with quantum computation, a hybrid approach combining CNN with certain phase error correction methods would produce the best solution.

Future work should focus on breaking the phase damping limit, hardware validation on real quantum devices, and scaling to more qubits. As quantum hardware capability increases, this approach can be extended to more complex quantum algorithms and error models, which could result in practical quantum advantage with fewer physical qubits than traditional error correction methods would.

## 11. Appendix:

1. GitHub repository link: https://github.com/KaranKendre11/MLforQuantumNoiseReduction/tree/main